# Lessons Learned from an Experiment in Crowdsourcing Complex Citizen Engineering Tasks with Amazon Mechanical Turk

MATTHEW STAFFELBACH, PETER SEMPOLINSKI, DAVID HACHEN, AHSAN KAREEM, TRACY KIJEWSKI-CORREA, DOUGLAS THAIN, DANIEL WEI AND GREGORY MADEY, University of Notre Dame

INTRODUCTION

1.1  Citizen Engineering

America's dated infrastructure is failing to keep pace with its burgeoning population. In fact, the average grade in ASCE's (American Society of Civil Engineers) 2013 report card for America's infrastructure was a D+, with a 3.6 trillion dollar estimated investment needed by 2020 and needs for inspection and assessment that far surpass available manpower. Crowdsourcing is increasingly being seen as one potentially powerful way of increasing the supply of labor for problem solving tasks, but there are a number of concerns over the quality of the data or analysis conducted. This is a significant concern when dealing with civil infrastructure for obvious reasons: flawed data could lead to loss of lives. Our goal was to determine if workers on Mechanical Turk were capable of developing basic engineering analysis skills using only the training afforded by comprehensive tutorials and guided questionnaires.

1.2  Citizen Science

Crowdsourcing has been effectively applied in the sciences, even prior to the internet. The Audubon society has been harnessing the power of the crowds in order to effectively plot the location of hundreds of bird species in the United States. Thousands of Audubon members would mail in information stating the number, species of birds, and their locations. Now the Audubon society and the Cornell lab of Ornithology run a real-time, online checklist program called eBird (ebird.org). Some other famous instances of effective citizen science include Galaxy Zoo (galaxyzoo.org), a galaxy classifying website and Phylo (Kawrykow et al. 2012) a game that allows crowds to help align related DNA sequences. Our goal was to test the possibility of Citizen Engineering for complex engineering tasks.

1.3  Amazon Mechanical Turk

We used Amazon Mechanical Turk (MTurk) as our crowdsourcing engine. MTurk has been shown to be a useful tool for conducting many types of research including behavioral research (Mason and Suri 2012, Crump et al. 2013), survey research (Buhrmester et al. 2011, Gosling et al. 2004), economic game experiments (Amir et al. 2012), collective behavior experiments (Suri and Watts 2011), and psychonomic research (Jasmin and Casasanto 2012). Others have shown that "turkers" (online workers from Amazon Mechanical Turk) can be used to ensure text translation quality (Marge et al. 2010) and twitter has used turkers to categorize trending tweets (Chen and Jain 2013).

Many of the typical uses for this website are image tagging, surveys, and audio transcription. In this paper we test the effectiveness of turkers at completing engineering tasks, building an army of so called "citizen engineers". Virtual Wind Tunnel (Bryson 1992) data analysis is used as a representative complex citizen-engineering task, because it would be unfamiliar to turkers and therefore would require some training (Sempolinski et al. 2012, Zhai et al. 2011). Also this particular task cannot be broken into many simple tasks "microtasks." Some other techniques for crowdsourcing complex tasks (Kittur et al. 2011) (Kulkarni 2012) would not be applicable, given that they concentrate on splitting the tasks into smaller non-complex tasks which do not require training. In this experiment every turker is required to read and comprehend a 4 to 5 page tutorial in order to begin to grasp the concepts necessary to effectively participate in any of these tasks.

## 2.0   Approach

We compared the skill of the anonymous turkers of Amazon Mechanical Turk, our unskilled crowd in assessing the quality of Virtual Wind Tunnel Data with the skill of two domain experts with formal training in the fields of fluid mechanics and fluid-structure interaction, who would serve as the source for ground truth. We released the 13 HITs (Human Intelligence Tasks) to two groups of turkers: turkers with 1) the masters qualification (a qualification awarded by Amazon) and 2) the default custom qualifications which requires the turkers to have completed at least 1000 HITs with a 95% approval rating. We also had two groups of graduate students with some coursework/training in fluid-structure interaction complete our HITs, students from the University of Notre Dame (USA) and Beijing Jiaotong University (PRC). These would be viewed as a skilled crowd for comparison sake. This first HIT contained the tutorial, a short survey and three simulation results that each had three graphical outputs to be assessed as indicators of simulation quality, as shown in Fig. 1. Each subsequent HIT contained three simulations; each simulation again contained three graphical outputs to be evaluated. This totaled to 117 questions for all the HITs not including the survey questions. 66 master workers completed our first HIT, 59 of them were qualified to move on the next HIT and 36 finished all the HITs. 51 Non-master turkers completed the 1st HIT, 27 were qualified to move on to the other HITs, and 9 finished all the HITs.

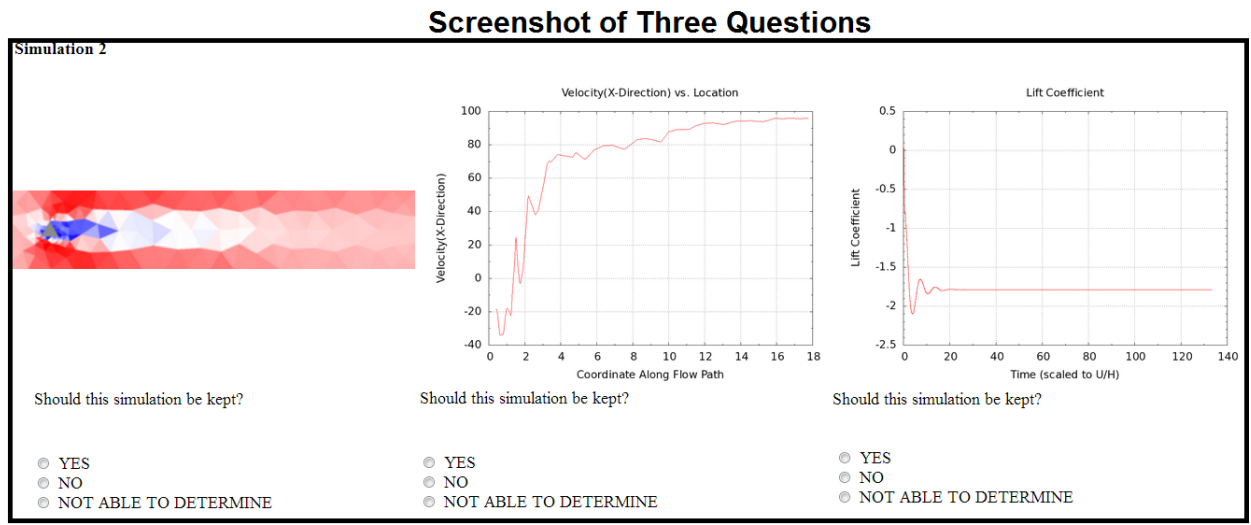

Fig. 1. This figure shows a screenshot of three questions that the turkers were asked to complete

## 2.1   Results

Our results (Fig. 2) showed that quality of the unskilled crowdworkers' work was slightly higher than that of the skilled crowd (graduate students). Along with showing that unskilled crowdworkers are effective at completing new complex tasks we also supply some explanations for this. Moreover, this paper provides number of important lessons that were learned in the process of collecting this data from MTurk, which should be of value to other researchers.

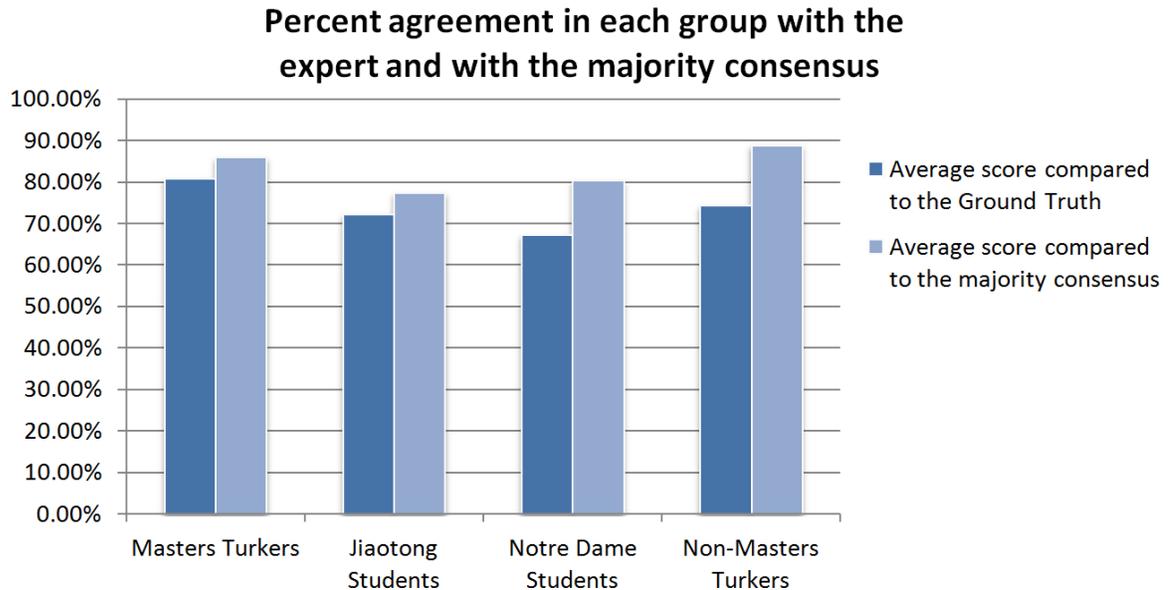

Fig. 2. Only turkers who completed all thirteen HITs with ten or less missing answers were included in these calculations. The most common answers of each group was calculated, this set of answers was denominated the majority consensus.

### 2.2   Who They Are

During this process of data collection we also studied the turkers' demographics by requiring that they complete a five-question survey. We discovered that 60% of turkers who chose to complete our first HIT had earned a college degree or higher and 71% (out of 39) of the turkers, who completed all 13 HITs had a college degree or higher. This may explain why they performed so well. In our survey we also asked, "[w]hich of the networks listed below have you ever used in order to enhance your use of Mechanical Turk (through discussions, ratings of requesters, finding new HITs, etc.)?" Only 13.5% of 59 turkers said that they did not use any networks; 60% said that they used MTurkForum on a regular basis. The other most popular sites were Turkopticon (turkopticon.ucsd.edu) and Turkernation (turkernation.com). All mentioned sites also include BlackHatWorld (blackhatworld.com), CloudMeBaby (cloudmebaby.com), Reddit(reddit.com), Facebook (facebook.com), LinkedIn (linkedin.com), and mTurk grind (mturkgrind.com). There was no evidence found that turkers were using these sites to "cheat" on our HITs. Instances were found where turkers would ask other turkers for clarification on how to approach some of the questions, but other turkers only responded by saying that the requester may not approve.

### 2.3   Lessons Learned

We spent over sixty hours reading through Mturk Forum and Turkernation threads, reviews, and general discussions. Mturk Forum and Turkernation have threads focused on great HITs, tips for avoiding scams, how to handle rejections, how much money you make in a day, and so forth, but there are a great number of other threads discussing TV shows, child names, computer gaming, and reading subjects. We also spent many hours on Turkernation's chat room. It was used mostly for notifying new HITs and discussing different requesters. Through this correspondence and personal messages sent to the requester a few notable ideas were gleaned. One is that there is a general distaste among the turker community for the use of the master qualification, mostly because many of them cannot figure out exactly how to become "qualified." More experienced turkers often become upset with newer turkers who will work for low paying requesters or those who are poor communicators. They are viewed in a similar way that union workers view those who work during strikes (scabs). The author of this paper found that many turkers were suspicious of new requesters and high paying HITs. Another lesson learned is that Turkopticon is a favored requester rating site for turkers. This site helps show what turkers believe is fair pay, how much communication is acceptable, the regular HIT acceptance speed, and the fairness of rejection rates.

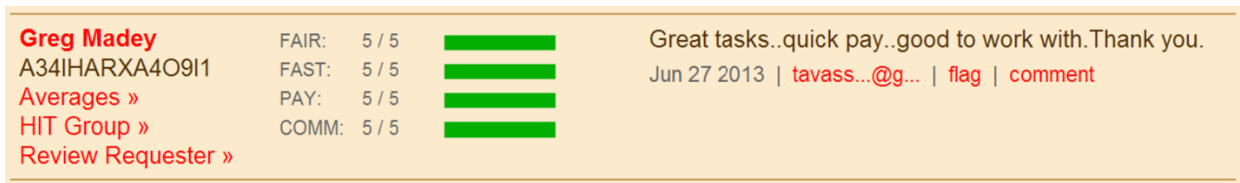

Fig. 3. This figure is a screenshot from the website Turkopticon that shows a sample review of the authors' requester profile

2.4 Suggestions

If a researcher is going to use Mturk, we suggest a few things:

1. Test your HIT by releasing a version where the description of the HIT is "test this HIT" and ask about 5 turkers to check your HIT for typos, and fill out all the answer boxes. Then check the data you get back to make sure the code functions properly.
2. Use Qualification HITs. This will be the first HIT you release. Only allow turkers who complete this HIT to move on to later HITs.
3. Use the Number of HITs qualification (>10,000 hits completed) and use the HIT approval rate qualification (>97% approved).
4. Or, use masters qualification, use it only on the qualification HIT then, after turkers are given your personalized qualification, you can take off the masters qualification knowing that all of your turkers are masters without paying the extra money that MTurk charges for using the masters qualification.
5. Make a page on Turkernation introducing yourself to turkers, and saying that you will bonus good workers and not reject bad workers (unless they are obviously scamming).
6. Watch your ratings on Turkopticon. If turkers are not accepting your HITs, it may be that some turkers have warned other turkers from working for you using Turkopticon.

3.0 Conclusions

Given a well-structured assessment process coupled with a comprehensive tutorial and compensation that averages around $6.50 an hour, turkers can and will complete long and complex engineering tasks with comparable competence to that of experts. If one is using the masters qualification the time it takes to complete your tasks is significantly increased. Master turkers are

significantly less likely to "spam" your HITs than non-masters turkers. Master turkers have a higher return rate than non-masters.

## Acknowledgements

The research presented was supported in part by an award from the National Science Foundation, under Grant No. CBET-0941565 for a project entitled *Open Sourcing the Design of Civil Infrastructure (OSD-CI), <http://www3.nd.edu/~opence>*. Dennis Staffelbach assisted in the editing of this paper. Zhi Zhai assisted in the construction of the HTML code used in the HITs.